\documentstyle[twoside]{article}

\catcode`\@=11
\long\def\@makefntext#1{
\protect\noindent \hbox to 3.2pt {\hskip-.9pt  
$^{{\eightrm\@thefnmark}}$\hfil}#1\hfill}		

\def\@makefnmark{\hbox to 0pt{$^{\@thefnmark}$\hss}}	
	
\def\ps@myheadings{\let\@mkboth\@gobbletwo
\def\@oddhead{\hbox{}
\rightmark\hfil\eightrm\thepage}   
\def\@oddfoot{}\def\@evenhead{\eightrm\thepage\hfil
\leftmark\hbox{}}\def\@evenfoot{}
\def\sectionmark##1{}\def\subsectionmark##1{}}

\oddsidemargin=\evensidemargin
\newcounter{sectionc}\newcounter{subsectionc}\newcounter{subsubsectionc}
\renewcommand{\section}[1] {\vspace{12pt}\addtocounter{sectionc}{1} 
\setcounter{subsectionc}{0}\setcounter{subsubsectionc}{0}\noindent 
	{\tenbf\thesectionc. #1}\par\vspace{5pt}}
\renewcommand{\subsection}[1] {\vspace{12pt}\addtocounter{subsectionc}{1} 
	\setcounter{subsubsectionc}{0}\noindent 
	{\bf\thesectionc.\thesubsectionc. {\kern1pt \bfit #1}}\par\vspace{5pt}}
\renewcommand{\subsubsection}[1] {\vspace{12pt}\addtocounter{subsubsectionc}{1}
	\noindent{\tenrm\thesectionc.\thesubsectionc.\thesubsubsectionc.
	{\kern1pt \tenit #1}}\par\vspace{5pt}}
\newcommand{\nonumsection}[1] {\vspace{12pt}\noindent{\tenbf #1}
	\par\vspace{5pt}}

\newcounter{appendixc}
\newcounter{subappendixc}[appendixc]
\newcounter{subsubappendixc}[subappendixc]
\renewcommand{\thesubappendixc}{\Alph{appendixc}.\arabic{subappendixc}}
\renewcommand{\thesubsubappendixc}
	{\Alph{appendixc}.\arabic{subappendixc}.\arabic{subsubappendixc}}

\renewcommand{\appendix}[1] {\vspace{12pt}
        \refstepcounter{appendixc}
        \setcounter{figure}{0}
        \setcounter{table}{0}
        \setcounter{lemma}{0}
        \setcounter{theorem}{0}
        \setcounter{corollary}{0}
        \setcounter{definition}{0}
        \setcounter{equation}{0}
        \renewcommand{\thefigure}{\Alph{appendixc}.\arabic{figure}}
        \renewcommand{\thetable}{\Alph{appendixc}.\arabic{table}}
        \renewcommand{\theappendixc}{\Alph{appendixc}}
        \renewcommand{\thelemma}{\Alph{appendixc}.\arabic{lemma}}
        \renewcommand{\thetheorem}{\Alph{appendixc}.\arabic{theorem}}
        \renewcommand{\thedefinition}{\Alph{appendixc}.\arabic{definition}}
        \renewcommand{\thecorollary}{\Alph{appendixc}.\arabic{corollary}}
        \renewcommand{\theequation}{\Alph{appendixc}.\arabic{equation}}
        \noindent{\tenbf Appendix \theappendixc #1}\par\vspace{5pt}}
\newcommand{\subappendix}[1] {\vspace{12pt}
        \refstepcounter{subappendixc}
        \noindent{\bf Appendix \thesubappendixc. {\kern1pt \bfit #1}}
	\par\vspace{5pt}}
\newcommand{\subsubappendix}[1] {\vspace{12pt}
        \refstepcounter{subsubappendixc}
        \noindent{\rm Appendix \thesubsubappendixc. {\kern1pt \tenit #1}}
	\par\vspace{5pt}}

\newcommand{\textlineskip}{\baselineskip=13pt}
\newcommand{\smalllineskip}{\baselineskip=10pt}

\def\eightcirc{
\begin{picture}(0,0)
\put(4.4,1.8){\circle{6.5}}
\end{picture}}
\def\eightcopyright{\eightcirc\kern2.7pt\hbox{\eightrm c}} 

\newcommand{\copyrightheading}[1]
	{\vspace*{-2.5cm}\smalllineskip{\flushleft
	{\footnotesize Modern Physics Letters A, #1}\\
	{\footnotesize $\eightcopyright$\, World Scientific Publishing
	 Company}\\
	 }}

\newcommand{\publisher}[2]{{\begin{center}\footnotesize\smalllineskip 
	Received #1\\
	\end{center}
	}}

\def\abstracts#1#2#3{{
	\centering{\begin{minipage}{4.5in}\baselineskip=10pt\footnotesize
	\parindent=0pt #1\par 
	\parindent=15pt #2\par
	\parindent=15pt #3
	\end{minipage}}\par}} 

\newcommand{\bibit}{\nineit}

\renewenvironment{thebibliography}[1]
	{\frenchspacing
	 \ninerm\baselineskip=11pt
	 \begin{list}{\arabic{enumi}.}
        {\usecounter{enumi}\setlength{\parsep}{0pt}     
	 \setlength{\leftmargin 12.7pt}{\rightmargin 0pt} 
         \setlength{\itemsep}{0pt} \settowidth
	{\labelwidth}{#1.}\sloppy}}{\end{list}}

\newcounter{itemlistc}
\newcounter{romanlistc}
\newcounter{alphlistc}
\newcounter{arabiclistc}

\newcommand{\fcaption}[1]{
        \refstepcounter{figure}
        \setbox\@tempboxa = \hbox{\footnotesize Fig.~\thefigure. #1}
        \ifdim \wd\@tempboxa > 5in
           {\begin{center}
        \parbox{5in}{\footnotesize\smalllineskip Fig.~\thefigure. #1}
            \end{center}}
        \else
             {\begin{center}
             {\footnotesize Fig.~\thefigure. #1}
              \end{center}}
        \fi}

\newcommand{\tcaption}[1]{
        \refstepcounter{table}
        \setbox\@tempboxa = \hbox{\footnotesize Table~\thetable. #1}
        \ifdim \wd\@tempboxa > 5in
           {\begin{center}
        \parbox{5in}{\footnotesize\smalllineskip Table~\thetable. #1}
            \end{center}}
        \else
             {\begin{center}
             {\footnotesize Table~\thetable. #1}
              \end{center}}
        \fi}

\def\@citex[#1]#2{\if@filesw\immediate\write\@auxout
	{\string\citation{#2}}\fi
\def\@citea{}\@cite{\@for\@citeb:=#2\do
	{\@citea\def\@citea{,}\@ifundefined
	{b@\@citeb}{{\bf ?}\@warning
	{Citation `\@citeb' on page \thepage \space undefined}}
	{\csname b@\@citeb\endcsname}}}{#1}}

\newif\if@cghi
\def\cite{\@cghitrue\@ifnextchar [{\@tempswatrue
	\@citex}{\@tempswafalse\@citex[]}}
\def\citelow{\@cghifalse\@ifnextchar [{\@tempswatrue
	\@citex}{\@tempswafalse\@citex[]}}
\def\@cite#1#2{{$\null^{#1}$\if@tempswa\typeout
	{IJCGA warning: optional citation argument 
	ignored: `#2'} \fi}}

\def\pmb#1{\setbox0=\hbox{#1}
	\kern-.025em\copy0\kern-\wd0
	\kern.05em\copy0\kern-\wd0
	\kern-.025em\raise.0433em\box0}

\def\fnt#1#2{\footnotetext{\kern-.3em
	{$^{\mbox{\scriptsize #1}}$}{#2}}}

\def\fpage#1{\begingroup
\voffset=.3in
\thispagestyle{empty}\begin{table}[b]\centerline{\footnotesize #1}
	\end{table}\endgroup}

\font\tenrm=cmr10
\font\tenit=cmti10 
\font\tenbf=cmbx10
\font\bfit=cmbxti10 at 10pt
\font\ninerm=cmr9
\font\nineit=cmti9

\font\eightrm=cmr8

\def\qed{\hbox{${\vcenter{\vbox{			
   \hrule height 0.4pt\hbox{\vrule width 0.4pt height 6pt
   \kern5pt\vrule width 0.4pt}\hrule height 0.4pt}}}$}}

\begin{document}

\normalsize\textlineskip
\thispagestyle{empty}
\setcounter{page}{1}

\copyrightheading{}			

\vspace*{0.88truein}

\fpage{1}
\centerline{\bf VACUUM FLUCTUATIONS AS A SOURCE OF }
\vspace*{0.035truein}
\centerline{\bf ADIABATIC PERTURBATIONS IN POWER LAW }
\vspace*{0.035truein}
\centerline{\bf INFLATIONARY COSMOLOGY }

\vspace*{0.37truein}
\centerline{\footnotesize Milan Miji\'c \footnote{ e-mail address: milan@moumee.calstatela.edu}}
\vspace*{0.015truein}
\centerline{\footnotesize\it Department of Physics and Astronomy, California State University}
\baselineskip=10pt
\centerline{\footnotesize\it Los Angeles, CA  90032 }
\vspace*{0.015truein}
\centerline{\footnotesize\it and Institute for Physics, P.O. Box 522}
\baselineskip=10pt
\centerline{\footnotesize\it 11001 Belgrade, Yugoslavia}
\vspace*{0.225truein}
\publisher{November 29, 1996}{}

\vspace*{0.21truein}
\abstracts{
An analysis of the decoherence of quantum fluctuations shows that production
of classical adiabatic density perturbations may not take place in models of
power-law inflation, $a(t) \sim t^p$, with $1< p < 3$. Some consequences
for models of extended inflation are pointed out. In general, the condition
for decoherence places new constraint on inflationary models, which does not
depend on often complicated subsequent evolution.}{}{}


\textlineskip			
\vspace*{12pt}			

\vspace*{-0.5pt}
\noindent
Among many attractive features of inflationary cosmology,$^{1,2}$
two stand out as perhaps the
deepest and the most successful ones: the solution to horizon problem,
and the mechanism for quantum origin of the primordial inhomogeneities.$^3$
At the same time, inflation is now in a peculiar state, rightly
characterized as a paradigm in search of the theory.$^2$ Perhaps
even more troubling is a persistent discrepancy  between the observed
low value of the density parameter and standard inflationary prediction of
$|\Omega - 1| \ll 1$.  

It seems reasonable in this situation to complement search for new
models with an examination of the stability and generality of those features of 
inflation which seem most desirable to persist in future cosmology. The 
mechanism for generation of primordial perturbations probably belongs to this 
category. 
While this kind of investigation is usually restricted to evaluation of
spectra, statistics, and overal normalization of the perturbations, we wish
to supplement such analysis with the examination of possible decoherence
of quantum fluctuations. That is, we would first like to see whether
vacuum fluctuations in a given cosmological model eventually do behave as 
(a source of) classical density perturbations.

This problem may be approached fairly generally.
Let $\phi$ stands for a homogeneous, minimally coupled scalar quantum field 
with potential $V(\phi)$, which drives a Robertson-Walker cosmological 
expansion with Hubble parameter $H$, while $\delta \phi (\vec {x}, t)$ 
denotes small, 
inhomogeneous fluctuations. The later quantity obeys the linearized field 
equation,$^4$

\begin{equation}
\delta \ddot \phi + 3 H \delta \dot \phi 
- a^{-2} \nabla^2 \delta \phi + V^{\prime \prime}(\phi) \delta \phi = 0
~~.
\label {linearized}
\end{equation}
\noindent
After expanding the field $\delta \phi$ into modes counted by the
wave number $k$,
introducing the conformal time $\eta$ as $dt = S(\eta) d\eta, ~~
S(\eta) = a(t)$, and the rescaled modes $\chi$ as $\delta \phi =
\chi /S(\eta) $, (the index $k$ will be suppressed), the one-mode equation
of motion becomes that for a time-dependent oscillator:

\begin{equation}
\chi^{\prime \prime}(\eta) + \omega^2 (\eta) \chi (\eta) = 0 ~~,
\label {osc}
\end{equation}
\begin{equation}
\omega^2 (\eta) = k^2 + m^2(\eta) S^2 - S^{\prime \prime}/S ~~.
\label {omega}
\end{equation}

\noindent
The time-dependent \lq\lq mass\rq\rq $~$ $m^2(\eta) \equiv 
V^{\prime \prime}(\phi)$ is to be evaluated from the solution of the
equation of motion for field $\phi$. Behavior of the frequency determines
the one-mode amplitude, from which one can calculate the occupation number
$n_k(t) \equiv \langle 0 | a_k^{\dagger}(t) a_k(t) | 0 \rangle $.
If, at late times (to be specified), this quantity becomes much greater than
unity, fluctuations of that comoving wavelength are effectively classical.$^5$
This is the approach that we will follow here. A very similar
physical picture, with somewhat different emphasis, has been thoroughly
developed in Ref. 6. For other approaches to the decoherence
of quantum fluctuations in inflationary universe, see Ref.'s 7 and 8.

In this paper we will extend the study of decoherence to quantum fluctuations 
in models of power law inflation.$^9$ It is a broad class of spatially flat, 
exact inflationary solutions with expansion
$a(t)= A t^p,~ p>1$, driven by an exponential potential,$^{10}$

\begin{equation}
V(\phi) = V_0~ e^{2 \phi/M} ~, ~~M\equiv \left ( \frac {p}{4\pi G} \right )^
{1/2}
~~.
\end{equation}
 
This situation occurs in several extensions of General Relativity, such as 
supergravity, Kaluza-Klein models, or Brans-Dicke gravity.$^2$
 
The analysis of decoherence for power law expansion turns out to be rather 
easy, due to the remarkable 
similarity with the massive De Sitter models, studied in Ref. 5.
In later case the evolution equation (\ref {osc}) is Bessel equation, with 
solutions given as $\chi (z) = \sqrt {z} B_{\nu} (z)$, $z \equiv - k \eta \in 
(0, \infty)$. $B_{\nu}$ is some Bessel function that satisfys chosen initial 
conditions. The late time ($z \rightarrow 0^+$) behavior of particle number 
was found to have the following simple form:$^5$
 
\medskip
{\it Particle number diverges when $\nu$ is real and different from $1/2$, 
and it is bounded, oscillatory function, always less than unity when $\nu$ is 
imaginary}.

\medskip  
Large occupation number in a given mode means that quasi-particles formed
a classical condensate. Thus, in terms of mass, we have that,

\medskip
{\it The vacuum fluctuations of massive fields on exact De Sitter background 
lead to classical density perturbations only for $m^2/H_0^2 < 2$, 
and $2 < m^2 \leq 9H_0^2/4$, while those of heavier fields do not}.

\medskip
$H_0$ stands for the constant Hubble parameter of the De Sitter space. The 
$m^2 = 2H_0^2$ case is exceptional, as it is conformally related to flat 
spacetime, so there is no particle production whatsoever.$^{11}$
While $m^2/H_0^2 \ll 2$ regime applies to standard slow-roll models,$^2$
the decoherence of heavy fields in a narrow range
$m^2/H_0^2 \in (2, 9/4]$, comes somewhat as surprise, since in this case
the potential is not of upside-down shape, and the standard explanation of
Guth and Pi$^7$ does not seem to apply. The physics of this case
is discussed in Ref. 12.

Returning to the power law inflation, we can express the scale factor in
the conformal time as $S(\eta) = B |\eta|^{p/(1-p)}, ~ \eta \in (- \infty, 0)$, 
and, $B \equiv A^{1/(1-p)} (p - 1)^{p/(1-p)}$. The one-mode frequency
(\ref {omega}) becomes,

\begin{equation}
\omega^2(\eta) = \alpha |\eta|^{-q} - \beta |\eta|^{-2} + k^2 ~~,
\end{equation}

\noindent
with $\alpha \equiv m^2 B^2 \geq 0$, 
$q \equiv 2p/(p-1)$, and $\beta \equiv p(2p-1)/(p-1)^2$.

The adiabatic perturbations are thought to originate in fluctuations of a 
field $\phi$ which drives the inflationary expansion.$^3$ One 
feature of such field in spatially flat power law expansion is that both 
kinetic and potential energy density vary as $1/t^2$.
This implies $m^2(t) \equiv \partial^2 V(\phi)/\partial \phi^2 
= 2 (3p -1)/t^2$.  After introducing the new variable $z \equiv - k \eta
\in (0, \infty)$, the expression for one-mode frequency (\ref {omega}) becomes,

\begin{equation}
\omega^2(z) = 1 + \frac {y(p)}{z^2}~,~~y(p) \equiv 
\frac {- 2p^2 + 7p -2}{(p-1)^2} ~~.
\label {omegay}
\end{equation}

Remarkably, the oscillator equation (\ref {osc}) again has the Bessel form,
so the one-mode amplitude will be given through Bessel function of the order
$\nu^2 = 1/4 - y(p)$. Since the late time behavior is again attained in the
$z \rightarrow 0^+$ limit, this case reduces to that of a massive field on De Sitter background, and we can apply the results obtained there. 
$\nu^2$ has zeros at $p=1/3$ and $p=3$, and it takes negative values between 
them. Therefore, we conclude that, 

\medskip
{\it The vacuum fluctuations of inflaton field which drives $a(t) \sim t^p$ 
power law inflation decohere only if $p \geq 3$}.

\medskip
 
For $1 < p < 3$, since $\nu$ is imaginary, the vacuum fluctuations do not
lead to generation of a classical (adiabatic) density perturbations,
despite the inflationary expansion. 

The corresponding two characteristic values are as follows: for $\nu = 0$, 
we have $m^2/H_0^2 = 9/4$ in De Sitter case, and $p = 3$ for case of power 
law inflation; $\nu = 1/2$ corresponds to $m^2/H_0^2 = 2$ in De Sitter
case, and $p = p_+ \equiv (7 + \sqrt {33})/4 \approx 3.186$, which is the
larger zero of the function $y(p)$ in the power law case.

The significance of the later value may be seen in several ways. First,
the relative change in one-mode frequency behaves as,

\begin{equation}
\left | \frac {\omega^{\prime}}{\omega} \right |=
\frac {|y(p)|}{z|z^2 + y(p)|} ~~.
\end{equation}

Denominator vanishes at $z_0 = |y(p)|^{1/2}$, providing that $y(p)$ takes
negative value. This can happen only for $p > p_+$, since the other zero
is at $p_-=(7 - \sqrt {33})/4 \approx 0.314 < 1$. This blow-up in 
$|\omega^{\prime}/\omega|$ leads to the divergence in the occupation number,
but as in De Sitter case it is due to the shrinking phase space 
($\omega^2(z_0) = 0$), rather than 
to large occupation number, and it has no physical significance. 
More importantly, for the same range of the parameter $p$, the oscillators turn 
upside down at that moment. And soon after, for {\it all} values of $p$, the 
change in frequency starts its rapid $1/z$ growth. Those two effects, working 
together, are responsible for the decoherence of fluctuations in $p > p_+$
models, when the oscillators are upside down. In a narrow range $p \in
(3, p_+)$ the oscillators are upside-right, and the fluctuations decohere
only because of the rapid change in frequency.$^{5,12}$

Further significance of the power $p=p_+$ may be seen from the fact that
the moment at which the wavelength of the mode equals the Hubble radius
is given as $z_{hc} = p/(p-1)$. As $p \rightarrow \infty$, the ratio 
$z_0/z_{hc}$ approaches the De Sitter value of $\sqrt 2$, and drops to 
zero as $p \rightarrow p_+$. This is because for $p<p_+$ the oscillator is
upside right, while just above that value the frequency starts its rapid growth 
only at asymptotically late times. The fluctuations for $p=p_+$
correspond to massless field conformally related to flat spacetime.
To see this, one may use Ref. 11 to evaluate the scalar curvature 
in power law models as
$R = 6p(2p-1)/t^2$. We have that $m^2(t) = R(t)/6$ holds only if $y(p)=0$.
Thus, in this case there
is no difference between the equation (\ref {linearized}) and that
for a massless, conformally coupled field, $(\partial^2  + R/6) \delta \phi 
= 0$.

[As a side remark, let us point out that results above do not depend on 
the usual ambiguity how 
we choose our quasi-particles. Let us choose some initial state, and let 
$(a, a^{\dagger})$ and $(b, b^{\dagger})$ denote annihilation and creation 
operators for two choices of quasi-particles. Both choices will be given as 
linear combinations of the one-mode amplitude and the associated momentum, 
but with a different coefficients. The two sets are related through the 
Bogoliubov coefficients $(\alpha_B, \beta_B)$, constrained as 
$|\alpha_B|^2 - |\beta_B|^2 = 1$ as both sets satisfy creation/annihilation
algebra. If, at any moment of time, we have a state with a definite number of
$a$ particles, the expectation value of $b$ particle number operator in that
state is given as,

\begin{equation}
n_b(z) = n_a(z) + |\beta_B|^2 ~~.
\end{equation}

The second term on the right hand side can not be negative.
Therefore, if $n_a(z)$ diverges as $z \rightarrow 0$, so does $n_b$.]

What we saw so far should be sufficient to illustrate that the study of
decoherence in inflationary cosmology does not merely confirm the
expectations, but leads to new restrictions on useful inflationary models.
Not every inflationary model leads to classical density perturbations. To
go one step further, let us briefly consider consequences of this analysis 
for a popular category of power law inflationary solutions, that of the 
so-called extended inflation.$^{13}$ 

Results obtained here should be applied to the picture in Einstein frame, 
where fluctuations can be cast in the form of equation (\ref {linearized}).
The power of the expansion $p$ is given through the Brans-Dicke field
self-coupling $b$ as, $p = b/2 + 3/4$. Production of density
perturbations in these models have been considered by several authors.$^{14}$
The amplitude of the inhomogeneity of a wavelength $\lambda$,
as it re-enters the Hubble radius in recent epoch, has been found to vary as 
$\lambda^{4/(2b-1)}$. A hope has been expressed that this departure from the 
scale-invariant spectrum may have some observational support. However,
none of these studies checked whether quantum fluctuations indeed decohere
in all of the cases. The classical nature of fluctuations was assumed in
analogy with the slow-rolling new inflationary models,$^3$ but
just as this thinking has been justified, for instance, by the analysis of
Guth and Pi,$^7$ a similar check should be done here. As we have
seen, for fluctuations in power-law models to be classical we must have 
$p \geq 3$. This implies that useful inflation must have Brans-Dicke self
coupling bounded as $b > 4.5$, and the enhancement towards longer wavelengths 
can not be faster than $\lambda^{1/2}$. 

The picture is even more dramatic if we consider the spectrum
of inhomogeneities, $P(k)$,  at some arbitrary (later) moment, when mode
is inside the Hubble radius. It follows
that $P(k) \sim k^{(2b-9)/(2b-1)}$, which translates into $P(k) \sim
k^{(p-3)/(p-1)}$. This approaches the Harrison-Zeldovich spectrum at
large $p$, but it only flattens out as $p$ decreases. The enhancement 
towards larger scales does not occur, as it corresponds to $p < 3$ 
expansion in which fluctuations do not decohere. And, in particular, some
of the versions of the extended inflation which rely on the slower 
expansion towards the end of that phase will not work. The constraint
$b > 4.5$ that follows from the decoherence constraint increases the lower
bound on Brans-Dicke coupling, usually thought to be bounded only as
$b > 1.5$.$^{13}$

A word of caution should be inserted here. The decoherence criteria that we
have used is expressed through the magnitude of the occupation number for a 
given mode. When this number is much larger than unity, inhomogeneity
of a given wavelength may be considered classical. This certainly looks
reasonable, and it is sufficient to explain the classical nature of
modes outside the Hubble radius in De Sitter universe.$^5$
It also agrees in its
physics with the criteria used by Guth and Pi,$^7$ or slightly
generalizes it.$^{12}$ However, all this is very different from other
mechanisms of decoherence,$^8$
which rely on the presence of interaction of fluctuations with some other
fields which are then traced out. This looks plausible, and similar 
to the decoherence processes in laboratory setting. The challenge of
that picture is that in each separate situation one must clearly state which
fields are to be kept track of, and which are to be summed over and why. 
It is possible
that the mechanism of this kind would lead to the classicality of 
fluctuations even in power law models with $p < 3$, but it is important to
bear in mind that this is something to be investigated, and not to be assumed
just because the expansion is inflationary. If such mechanism exists, the
rate of decoherence will depend on some coupling constant, which may lead
to additional constraints on the model.

To conclude, let us recall that there are many inflationary models that pass 
various criteria, and it is still too early to say whether any of them has 
particular advantage over the others. Placing constraints on a model from
comparison with the present day universe is mandatory, but tricky business.
For example, the use of fluctuations for the 
formation of large scale structure depends on a number of 
details in a specific cosmological scenario. Most of the recent models use 
more than one scalar field, and the evolution is rather complicated.
In contrast, the 
decoherence analysis done here produces criteria for successful inflationary 
cosmology which is based on the requirement for the very existence of
classical inhomogeneities. As such, this statement is very robust, as
it does not depend on details of the subsequent evolution.
The method that we have used here may in principle be applied
to any other, exact or approximate, scalar field driven cosmological 
model.

\nonumsection{References}


\begin{thebibliography}{999}
\bibitem{1} A. Guth, {\bibit Phys. Rev. D23} (1981) 347.
\bibitem{2} E. Kolb and M.S. Turner, 
{\bibit The Early Universe}, (Addison Wesley, 1990); 
A. Linde, {\bibit Particle Physics and Inflationary Cosmology},
(Harwood Academic Press, 1990). 
\bibitem{3} S.W. Hawking, {\bibit Phys. Lett. 115B} (1982) 295;
A.A. Starobinsky, {\bibit Phys. Lett. 117B} (1982) 175;
A. Guth and S-Y. Pi,{\bibit Phys. Rev. Lett.  49} (1982) 1110; 
J.M. Bardeen, P.J. Steinhardt, and M.S. Turner, {\bibit Phys. Rev. D28}
(1983) 679.
\bibitem{4} A. Guth and S-Y. Pi, Ref. 2.
\bibitem{5} M. Miji\'c, Cal State LA preprint, August 1996 (submitted).
\bibitem{6} D. Polarski and A.A. Starobinsky, {\bibit Class. Quan. Grav. 13}
(1996) 377; J. Lesgourgues, D. Polarski, and A.A. Starobinsky, preprint
gr-qc/9611019.
\bibitem{7} A. Guth and S-Y. Pi, {\bibit Phys. Rev. D32} (1985) 1899.
\bibitem{8} M. Sakagami, {\bibit Prog. Theor. Phys. 79} (1985) 443;
R. Brandenberger, R. Laflamme, and M. Miji\'c, {\bibit Mod. Phys. Lett. A 28}
(1990) 2311; 
Y. Nambu, {\bibit Phys. Lett. 276B} (1992) 11;
B.L. Hu, J.P. Paz, and Y. Zhang, in {\bibit The
Origin of Structure in the Universe}, ed. E. Gunzig and P. Nardone, (Kluwer, 
1993).
\bibitem{9} L. Abbott and M. Wise, {\bibit Nucl. Phys. B224}
(1984) 541; F. Lucchin and S. Mataresse, {\bibit Phys. Rev. D32}
(1985) 1316.
\bibitem{10} G.F.R. Ellis and M. Madsen, {\bibit Class. Quan. Grav. 8} (1991) 
667; F. Lucchin and S. Mataresse, Ref. 9.
\bibitem{11} N.D. Birrell and P.C.W. Davies, {\bibit Quantum
Fields in Curved Space}, (Cambridge University Press 1982).
\bibitem{12} M. Miji\'c, Cal State LA preprint, December 1996 (submitted).
\bibitem{13} D. La and P. Steinhardt {\bibit Phys. Rev. Lett. 62}
(1989) 376; {\bibit ibid}, {\bibit Phys. Lett. 220B} (1989) 375; 
D. La, P. Steinhardt, and E. Bertshinger, {\bibit Phys. Lett. 231B} (1989) 231.
\bibitem{14} E. Kolb, D. Salopek, and M.S. Turner, {\bibit Phys. Rev.
D42} (1990) 3925; A.H. Guth and B. Jain, {\bibit Phys. Rev. D45} 
(1992) 426; R. Crittenden and P.J. Steinhardt, {\bibit Phys. Lett. 293B}
(1992) 32.

\end{thebibliography}
\end{document}